\documentclass[final]{svjour2}
\usepackage{graphicx}
\usepackage{rotating}
\usepackage{amssymb}
\usepackage{amsmath}
\usepackage{mathptmx}
\usepackage[numbers]{natbib}
\usepackage{color}
\usepackage{macros}
\makeatletter
\journalname{Journal of Low Temperature Physics}
\usepackage[pdftex]{hyperref}
\begin{document}
\title{Superfluid phases of \He\  in a periodic confined geometry\footnote{\small Published online in J. Low Temp. Phys. (2013) 
[\href{http://dx.doi.org/10.1007/s10909-013-0924-4}{doi: 10.1007/s10909-013-0924-4}].}
}
\author{J. J. Wiman \and J. A. Sauls}
\institute{Department of Physics \& Astronomy, Northwestern University, Evanston, IL 60208 USA\\
\email{jjwiman@u.northwestern.edu}}
\date{\today}
\maketitle

\begin{abstract}
Predictions and discoveries of new phases of superfluid \He\ in confined geometries, as well as
novel topological excitations confined to surfaces and edges of near a bounding surface of \He, are
driving the fields of superfluid \He\ infused into porous media, as well as the fabrication of
sub-micron to nano-scale devices for controlled studies of quantum fluids.
In this report we consider superfluid \He\ confined in a periodic geometry, specifically a
two-dimensional lattice of square, sub-micron-scale boundaries (``posts'') with translational
invariance in the third dimension.
The equilibrium phase(s) are inhomogeneous and depend on the microscopic boundary conditions imposed
by a periodic array of posts. We present results for the order parameter and phase diagram based on
strong pair breaking at the boundaries. The ordered phases are obtained by numerically minimizing
the Ginzburg-Landau free energy functional. We report results for the weak-coupling limit,
appropriate at ambient pressure, as a function of temperature $T$, lattice spacing $L$, and post
edge dimension, $d$.
For all $d$ in which a superfluid transition occurs, we find a transition from the normal state to a
periodic, inhomogeneous ``polar'' phase with $T_{c_1} < T_{c}$ for bulk superfluid \He.
For fixed lattice spacing, $L$, there is a critical post dimension, $d_c$, above which
only the periodic polar phase is stable.
For $d < d_c$ we find a second, low-temperature phase onsetting at $T_{c_2} < T_{c}$ from the polar
phase to a periodic ``B-like'' phase. The low temperature phase is inhomogeneous, anisotropic and
preserves time-reversal symmetry, but unlike the bulk B-phase has only $\DfourhLS$ point symmetry.
\par
\smallskip
\end{abstract}

\section{Introduction}

The p-wave, spin-triplet superfluid phases of \He\ provide the paradigm for unconventional BCS
pairing in which spin- and orbital rotation symmetries, $\spin\times\orbital$, are spontaneously
broken in conjunction with $\gauge$ gauge symmetry.
It was realized soon after the discovery that these broken symmetries, particularly parity and
orbital rotation symmetry, implied that interfaces, boundaries and impurities could have profound
effects on the superfluid phases.\cite{amb75,rai77}
In the case of the bulk A-phase the effect of the boundary is to \emph{lock} the orbital
quantization axis, $\hat{l}$, normal to the boundary.
The influence of boundaries can often extend to length scales much longer than the coherence length,
$\xi_{0}=\hbar v_{f}/2 \pi k_{\text{B}} T_{c}\approx 200-800\,\mbox{\AA}$ depending on pressure,
when there is competition between alignment effects from curved boundaries and/or
superflow.\cite{deg74}
In a long cylinder with radius $R\gg\xi_{0}$ the boundary condition on $\hat{l}$ leads to a
\emph{texture}, i.e. a long-wavelength spatial variation of the orbital quantization axis,
$\hat{l}$, which is also an equilibrium current-carrying state.\cite{mer76,wal12,kun13}
At the coherence length scale near a boundary strong pair-breaking typically occurs. The orbital
component of the order parameter normal to the surface is suppressed and a spectrum of Fermionic
states are localized near the boundary.\cite{buc81,nag98,vor03} The de-pairing effect of the
boundary is further enhanced if the surface is rough on length scales comparable to or smaller than
$\xi_{0}$.
If superfluid \He\ is confined to a region with dimensions of order a few coherence lengths then the
geometry and surface structure on the boundaries can significantly modify the equilibrium phase
diagram, and can even stabilize phases not realized in bulk superfluid
\He.\cite{kja78,nag98,vor03,vor07}

Several studies of superfluid \He\ have been performed on thin films or within a slab
geometry.\cite{fre88,xu90,sch98} In the case of strong one-dimensional confinement, i.e. boundary
separation $D < D_{c_2}\approx 9\,\xi_{0}$, the A phase is expected to be the stable phase even at
pressures well below the bulk critical pressure, $p_c\approx 21\,\mbox{bar}$.\cite{vor03}
NMR measurements strongly support this prediction.\cite{lev13} This is in stark contrast to the bulk
\He\ phase diagram in which the A phase is only stable at high temperature and pressure.
In weak-coupling theory the planar and axial (ABM) phases are degenerate even with strong surface
disorder.\cite{vor03} Strong-coupling effects which stabilize the ABM state at high pressures and
high temperatures are poorly known for inhomogeneous phases at low-temperatures, $T\ll T_c$, and low
pressures, $p\rightarrow 0\,\mbox{bar}$. For this reason the ground state of thin \He\ films at the
lowest pressures is still an open question.
At intermediate scales of confinement, $D_{c_2} < D < D_{c_1}\approx 13\,\xi_{0}$, the ground state
in the weak-coupling limit is predicted to be a ``crystalline'' phase with an order parameter that
spontaneously breaks translation symmetry in the plane of the film.\cite{vor07} A one-dimensional
periodic phase (``striped phase'') with in-plane wavelength $Q_{\perp}^{-1}\approx 3\,\xi_{0}$ has
lower energy than any of the translationally invariant axial, planar or B-planar phases over a wide
range of film thicknesses and temperatures. The mechanism responsible for spontaneously breaking
translation symmetry for $D\lesssim D_{c_1}$ is the energy cost of surface pair-breaking compared to
the energy cost for domain wall formation between degenerate B-planar phases. For $D < D_{c_1}$ it
is energetically favorable for domain walls to enter the film. Interactions between domain walls
lead to the striped phase.
This type of competition between surface pair-breaking, the formation of topological defects and the
stabilization of new phases not realized in bulk \He\ is part of the motivation for developing
sub-micron to nano-scale geometries for confining \He.\cite{lev13,gon13} Of particular interest for
this study is the possibility of confining \He\ in a periodic geometry such as a cavity supported by
a periodic array of sub-micron scale posts.\cite{zhe13}

We break translational symmetry externally by considering \He\ infused into an infinite
two-dimensional ($x-y$) periodic array of square posts, with translational invariance in the third
dimension ($z$).
This geometry can also be viewed as a two-dimensional ($2d$) grid formed of vertical ($x$) and
horizontal ($y$) channels. The spatial region between the corners of four adjacent posts, or
alternatively where the $x$ and $y$ channels intersect is particularly significant, and we will
refer to this region as the ``center'' of the $2d$ cell when discussing superfluid \He\ confined
within this structure.
We expect the results reported here to be valid for confinement lengths in the $z$ dimension,
$L_z\gg 30\xi_0$.

The order parameter for superfluid \He\ belongs to the manifold of spin-triplet, p-wave, BCS pairing
states represented by the $2\times 2$ ``gap matrix'',
\be
\hDelta(\hat{p}) = \sum_{\alpha i}(i \vec{\sigma}_{\alpha}\sigma_{y})\,A_{\alpha i}\,\hat{p}_{i}
\,,
\ee
which is a function of the direction of relative momentum of the Cooper pair, $\hat{p}$, and is
parametrized in its most general form by nine complex amplitudes, $A_{\alpha i}$.
The $3\times 3$ matrix order parameter transforms as a vector under spin rotations, and separately
as a vector under orbital rotations. The maximal symmetry group for bulk \He\ is ${\mathsf G} =
\gauge\times\spin\times\orbital\times\parity\times\time$, where $\parity$, $\time$ and $\gauge$
represent space inversion, time-reversal and global gauge symmetries of the normal phase.
The symmetry reduction resulting from the weak nuclear dipolar interaction is omitted here, but is
important in resolving relative spin-orbit rotational degeneracies, and in determining the NMR
signatures of the phases of confined \He.

\section{Ginzburg-Landau Theory}

To determine the phase diagram and superfluid order parameter for \He\ confined within a $2d$
periodic structure we minimize the Ginzburg-Landau (GL) free energy for a general spin-triplet,
p-wave condensate defined as a functional of the $3\times 3$ matrix order parameter.
A few atomic units away from a boundary the \He-\He\ interactions responsible for pairing are
invariant under the maximal symmetry group of \underline{bulk} \He. Thus, the GL functional takes
its bulk form,\cite{rai76,vollhardt90}
\begin{align}\label{eq:glfe}
\Omega[A] 
&=\int_{V}d\vec{R}\;
\left\{
         \alpha(T) Tr\left(A A^{\dagger}\right) 
        +\beta_{1} \left|Tr(A A^{T})\right|^{2}
		+\beta_{2} \left[Tr(A A^{\dagger})\right]^{2}  
		\right.
\nonumber  \\
& \qquad \left.
		\vphantom{\frac{1}{3}} 
		+\beta_{3}\, Tr\left[A A^{T} (A A^{T})^{*}\right] 
		+\beta_{4}\, Tr\left[(A A^{\dagger})^{2}\right] 
		+\beta_{5}\, Tr\left[A A^{\dagger} (A A^{\dagger})^{*}\right] 
		\right.
\\
& \qquad \left.
		\vphantom{\frac{1}{3}} 
		+K_{1} \left(\nabla_{k}A_{\alpha j} \nabla_{k}A_{\alpha j}^{*}\right)
		+K_{2} \left(\nabla_{j}A_{\alpha j} \nabla_{k}A_{\alpha k}^{*}\right)
		+K_{3} \left(\nabla_{k}A_{\alpha j} \nabla_{j}A_{\alpha k}^{*}\right) 
\right\}
\,.
\nonumber
\end{align}

The equilibrium order parameter is obtained from the stationarity condition for the GL functional. 
Confinement is introduced via boundary conditions of the order parameter field, 
$A_{\alpha i}(\vec{R})$.
For \He\ confined in a non-magnetic, periodic geometry with $4$-fold rotational, reflection and
inversion symmetries, the maximal symmetry group is reduced by restricting the orbital rotations to
the point group $\Dfourh$ (which includes space inversion), i.e.
\be
{\mathsf G} = \gauge\times\spin\times\Dfourh\times\time
\,.
\ee
The domain, $V$, is a square unit cell with side length $L$. Periodic boundary conditions are
imposed on the order parameter field at the outer boundaries of this unit cell. In the interior of
the unit cell is an inner boundary representing the square post of side length $d$. 
Typical boundary conditions for the order parameter on the inner boundary are: (i) \emph{maximal}
pair-breaking in which all components of the order parameter vanish on the inner boundary and (ii)
\emph{minimal} pair-breaking in which only the orbital component normal to the surface of the inner
boundary is forced to vanish, and the normal derivative of the tangential orbital components
vanishes on the inner boundary. This latter boundary condition corresponds to surfaces with specular
reflection,\cite{amb75} while the former boundary condition corresponds to an atomically rough
surface with strong backscattering.\cite{sau11} Here we report results based on maximal 
pair-breaking.
We numerically minimize the GL functional on this domain, and determine the stable (and in some
cases meta-stable) order parameter (phases) for superfluid \He\ in this class of periodic confined
geometries.
We also present results for the phase diagram as a function of temperature $T$, confinement length
$D\equiv L-d$, and period $L$. The results reported here are appropriate for low pressures in the GL
regime. Thus, we assume weak-coupling values for the GL material parameters:\cite{rai76,vollhardt90}
\begin{align}
\alpha(T) & = \frac{1}{3}N(0)(T/T_{c}-1)
\,,\qquad
%
2\beta_{1} = -\beta_{2} = -\beta_{3} = -\beta_{4} = \beta_{5}
\,,
\nonumber \\
K_{1} & = K_{2} = K_{3} = \frac{3}{5} N(0) \xi_{0}^2
\,,\quad
\beta_{1}=- \frac{N(0)}{(\pi k_{\text{B}}T_{c})^2}
            \left\{\frac{1}{30}\left[\frac{7}{8}\zeta(3)\right]\right\}
\,,
\end{align}
where $T_c$ is the superfluid transition temperature for bulk \He, $\xi_{0}$ is the zero-temperature
correlation length, and $N(0)=k_{f}^{3}/2\pi^2\,v_{f} p_{f}$ is the single-spin quasiparticle
density of states at the Fermi surface, defined in terms of the Fermi velocity, $v_f$, and Fermi
momentum and wavenumbers, $p_f=\hbar k_f$.

In what follows we neglect the nuclear dipolar energy and choose \emph{aligned} spin and orbital
coordinate axes, $\{x,y,z\}$, corresponding to the high symmetry directions of the periodic channel.
Thus, the order parameter is represented by,
\begin{align}
A &= 
\begin{pmatrix} 
A_{xx} & A_{xy} & A_{xz} \\
A_{yx} & A_{yy} & A_{yz} \\
A_{zx} & A_{zy} & A_{zz} 
\end{pmatrix}
\,.
\end{align}
For bulk \He, the B-phase, defined by the Balian-Werthamer state,
\be
A^{\text{B}} = \frac{\Delta_{\text{B}}}{\rootthree}\,
               \begin{pmatrix} 
					1 & 0 & 0 \\
					0 & 1 & 0 \\
					0 & 0 & 1 
				\end{pmatrix}
\,,
\ee
with amplitude given by
\be
\Delta_{\text{B}}^2(T) = \frac{1}{2}\frac{|\alpha(T)|}{\beta_{12}+\frac{1}{3}\beta_{345}}
\,,
\ee
is the equilibrium phase at low pressures with free energy density given by
\be
\Omega_{\text{B}}/V 
= -\frac{1}{4}\frac{|\alpha(T)|^2}{\beta_{12}+\frac{1}{3}\beta_{345}}
= -\frac{\Delta C_{\text{B}}}{2T_{c}}\,(T-T_{c})^2
\,.
\ee
The second term is the B-phase condensation energy scaled in terms of the heat capacity jump,
$\Delta C_{\text{B}}$, at the normal to B-phase transition. For weak-coupling values of the material
coefficients this gives the BCS result, $\Delta C_{\text{B}}/C_{\text{N}} = 12/7\zeta(3)\simeq
1.43$, where $C_{\text{N}}=\twothirds \pi^2 N(0)\,T_c$ is the normal-state heat capacity at $T_c$.
These values for the bulk B-phase order parameter and free energy are used as the scale for the
order parameter and free energy of confined \He.

We note that the boundary condition imposed by the interior post is expected to be accurate only for
post side lengths $d \gtrsim \xi_{0}$. For post dimensions, $d \ll \xi_{0}$, the boundary is more
accurately treated microscopically as an ``impurity'' that scatters excitations and breaks
pairs.\cite{rai77} The pair-breaking effect of an impurity with side dimension smaller than a
coherence length is reduced by $d/\xi_{0}$ near the post. We avoid this limit and restrict our
analysis to post dimensions with $d \ge \frac{1}{2} \xi_{0}$.

Before discussing the numerical results we describe some of the possible phases with a high degree
of residual symmetry, i.e. sub-groups of the maximal symmetry group, that may be realized by \He\ in
a confined $\Dfourh$ geometry.

\section{Symmetry Classes of \He\ in a confined $\Dfourh$ geometry}\label{Sec-Symmetry_Classes}

The effects of confinement are enforced by the boundary conditions imposed on the order parameter.
The boundary conditions reflect the point symmetry of the confining boundaries.
For the case of a periodic array of square posts the elementary symmetry group of a square post,
$\CfourV$, is the combined set of four-fold rotations, $\{E, C_4, C_4^2, C_4^3\}$, where $E$ is the
identity and $C_4^n = (C_4)^n$ is a rotation about the $z$ axis by $n\times\pi/2$, the set of
reflections through four vertical planes, $\{\Pi_{zx}, \Pi_{zy}, \Pi_{zx'}, \Pi_{zy'}\}$, and the
corresponding rotary reflections, $\{R_{zi}\equiv C_4\Pi_{zi}\,|\, i=x,y,x',y'\}$, where $(x',y')$
are axes rotated from $(x,y)$ by $\pi/4$ about $z$.
The addition of reflection symmetry through the horizontal plane, $\Pi_{xy}$, and $180^{\circ}$
rotations about the vertical plane symmetry axes, $\{C_{2x}, C_{2y}, C_{2x'}, C_{2y'}\}$, defines
the point group, $\Dfourh$, which includes space inversion, $C_{i}=C_{2y}\cdot\Pi_{zx}$.

For any element $g\in\Dfourh$ a scalar function transforms as
$f(\vec{R})\xrightarrow[]{g}\,f(\hat{g}^{T}\cdot\vec{R})$, where $\hat{g}$ is the $3\times 3$
matrix representing the symmetry element $g$, $\hat{g}^{T}=\hat{g}^{-1}$ is the matrix inverse, and
$\vec{R}=(x,y)$.
Thus, the order parameter field, which is a vector under space rotations and reflections, transforms
as $A_{\alpha i}(\vec{R})\xrightarrow[]{g}g_{ij}A_{\alpha j}(\hat{g}^{T}\cdot\vec{R})$.
Similarly, for any rotation $g\in\spin$ we have, $A_{\alpha i}(\vec{R})\xrightarrow[]{g}
g_{\alpha\beta}A_{\beta i}(\vec{R})$, and under a gauge transformation, $\chi\in\gauge$, $A_{\alpha
i}(\vec{R})\xrightarrow[]{\chi}\,e^{-i\chi}\,A_{\alpha i}(\vec{R})$. Time-reversal, $\time$, reduces
to complex conjugation, $A_{\alpha i}(\vec{R})\xrightarrow[]{\time}\,A_{\alpha i}^{*}(\vec{R})$.

\subsection{Non-Equal Spin Pairing - The $\Bsquare$ Phase}

In the case of bulk \He\ the maximal symmetry sub-group of joint spin and orbital rotations combined
with time-reversal, $\spinorbital\times\time$, is the symmetry class of the B-phase, i.e. the
Balian-Werthamer state with $A_{\alpha i}^{\text{BW}}=\Delta\,\delta_{\alpha i}$. The \emph{discrete
analog} of the bulk B-phase is a state, which we refer to as the $\Bsquare$-phase, that is invariant
under \underline{joint} spin and orbital elements of the maximal point group, $\Dfourh$, and
time-reversal, i.e. $H_{\Bsquare} = \DfourhLS\times\time$. Note that space inversion is broken, but
space inversion combined with inversion in spin-space is a symmetry of the $\Bsquare$-phase.

The matrix structure of the $\Bsquare$ order parameter differs substantially from the isotropic
$B$-phase. For all $g\in H_{\Bsquare}$, the order parameter satisfies,
\be
A_{\alpha i}(\vec{R})
\xrightarrow[]{g} g_{\alpha\beta}\,A_{\beta j}(\hat{g}^{T}\cdot\vec{R})\,g^{T}_{ji} 
= A_{\alpha i}(\vec{R})
\,.
\ee
It is then straightforward to show that
\be
A^{\Bsquare} = \begin{pmatrix} A_{xx}	& A_{xy} 	&	0 \\
							   A_{yx}	& A_{yy}	&	0 \\
										0				&				& A_{zz} \end{pmatrix} 
\,,
\ee
with $A_{xx}(x,y)=A_{yy}(y,x)$, $A_{xy}(x,y)=A_{yx}(y,x)$, $A_{zz}(x,y)=A_{zz}(y,x)$ and all
components are real ($\time$ symmetry). The diagonal (off-diagonal) components are even (odd) under
$x\rightarrow -x$ or $y\rightarrow -y$.
The numerical results presented below show that the $\Bsquare$-phase is the equilibrium state in the
weak-coupling limit at low temperatures.

\subsection{Equal Spin Pairing States}

The superfluid phases with the highest degree of residual symmetry are those that preserve a
continuous rotation symmetry about an axis $\hat{d}$ in spin space, i.e. $\twoDspin$. The direction
$\hat{d}$ is a vector representing spontaneously broken spin-rotation symmetry. If $\hat{d}$ is
real, then the broken symmetry phase is an equal-spin-pairing (ESP) state and $\hat{d}$ is the
direction in which the Cooper pairs have zero spin projection. The residual symmetry group of the
class of ESP states includes $\twoDspin\times\ZtwoSpin$, where $\ZtwoSpin=\{1,
e^{i\pi}\,R^{\text{spin}}_{\pi\hat{x}}\}$ is a two element group of the identity and the combined
operation of a gauge transformation, $e^{i\pi}$, and a rotation of $\pi$ about an axis
$\hat{x}\perp\hat{d}$ in spin space. Continuous $\gauge$ symmetry is broken, but elements of
$\gauge$ may be combined with spin or orbital rotations and reflections.
In particular, the point symmetry, $\Dfourh$, is necessarily broken by any p-wave pairing state.
However, the residual symmetry of the z-aligned polar ($P_z$) phase contains all the elements of
$\Dfourh$. In particular, the $P_z$ phase can be expressed as $A_{\alpha
i}^{P_z}=\hat{d}_{\alpha}\,a(x,y)\,\hat{z}_i$ with $a(x,y)$ real (time-reversal symmetry) and
invariant under the sub-group $\CfourV$: $a(x,y)=a(y,x)=a(-x,y)=a(x,-y)$. The $P_z$ order parameter
undergoes a sign change for any of the $C_{2i}$ operations and reflection in the $xy$ plane since
$\hat{z}\rightarrow -\hat{z}$. Thus, combining these operations with the gauge transformation,
$e^{i\pi}$, and $\CfourV$ yields the group, $\DfourhLpi$, which is isomorphic to $\Dfourh$. Thus,
the residual symmetry group for the $P_z$ phase is
\be
{\mathsf H}_{P_z} = \twoDspin\times\ZtwoSpin\times\DfourhLpi\times\time
\,.
\ee
This state is the stable superfluid phase that onsets from the normal state at $T_{c_1}$.

The $P_z$ phase retains the sub-group $\CfourV$ of point symmetries, but is not the only ESP state
with this symmetry. If we omit the operations that transform $z\rightarrow -z$, then we obtain two
possible symmetry classes. If time-reversal is preserved we obtain the residual symmetry group is
\be
{\mathsf H}_{\text{3D}} = \twoDspin\times\ZtwoSpin\times\CfourVL\times\time
\ee
with an orbital order parameter field, 
$\vec{a} = a_x \hat{x} + a_y \hat{y} +a_z \hat{z}$, that includes three real components
satisfying the reflection symmetries, 
\ber
a_x(x,y) &=& -a_x(-x,y) = +a_x(x,-y)
\label{eq-C4v_reflections_ax}\\
a_y(x,y) &=& +a_y(-x,y) = -a_y(x,-y)
\label{eq-C4v_reflections_ay}\\
a_x(x,y) &=& a_y(y,x) 
\label{eq-C4v_reflections_axy}\\
a_z(x,y) &=& a_z(y,x) = a_z(-x,y) = a_z(x,-y)
\,.
\label{eq-C4v_reflections_az} 
\eer
This phase is not found to be a local minimum of the GL functional for the weak-coupling values of
the $\beta$ parameters.

Another ESP phase with $\CfourVL$ symmetry is obtained if we break $\time$ symmetry, but preserve
$\Pi_{xy}\cdot\time$. In this case the residual symmetry group is
\be
{\mathsf H}_{\text{chiral}-\CfourV} 
= \twoDspin\times\ZtwoSpin\times\CfourVL\times\{E\,,\,\Pi_{xy}\cdot\time\}
\,.
\ee
The orbital vector, $\vec{a}_{\pm} = \vec{a}_{\perp} \pm i a_z \hat{z}$, with 
$\vec{a}_{\perp} = a_x \hat{x} + a_y \hat{y}$, is a complex vector field with
real amplitudes $\{a_x,a_y,a_z\}$ satisfying the reflection symmetries in Eqs.
\ref{eq-C4v_reflections_ax}-\ref{eq-C4v_reflections_az} required by $\CfourVL$.
The $\pm$ sign reflects the two-fold degeneracy resulting from broken time-reversal symmetry.
These are \emph{chiral} phases with a local chiral vector field given by
\be
\vec{l}_{\pm} = \pm\,\vec{a}_{\perp}\times\,a_z\hat{z} 
              = \pm\,a_z(\vec{R})\left[a_y(\vec{R})\,\hat{x} - a_x(\vec{R})\,\hat{y}\right]
\,,
\ee
which is confined to the $xy$ plane. For a unit cell centered on the post, the chiral vector
vanishes on the post boundaries \underline{and} at the center of the two channels, where
$a_x=a_y=0$, \underline{if} the periodicity of the ordered phase is the same as that of the
underlying geometry. Here the phase is locally the $P_z$ phase. For chiral phases, and more
generally current carrying states, the periodicity of the ordered phase need not equal the
underlying lattice periodicity. Thus, a complete classification of the residual symmetry sub-groups
should include the space-group operations. This is beyond the scope of this report, but underscores
the complexity of the possible phases of \He\ in a periodically confined geometry.
In the weak-coupling limit this chiral ESP phase is not energetically stable, but this phase, or a
closely related phase with period $2L$, may emerge as a stable, or meta-stable, low temperature
phase at high pressures due to strong coupling effects.\footnote{The in-plane chiral phase with
period $2L$ is a periodic version of the texture obtained by Surovtsev and Fomin \cite{sur08} for
a uniform distribution of rod-like impurities embedded in \Hea.}
However at high pressures, for very weak confinement, $L\gg 20\xi_0$, and small post dimensions,
$d\lesssim\xi_0$, the $\mathsf{chiral}-\CfourVL$ phase is unlikely to be the equilibrium phase. In
this limit we expect a chiral ABM-like phase with $\vec{l}||\hat{z}$ in the center of the channels
to be the equilibrium phase at temperatures below a narrow region of stability of the $P_z$ phase.

The residual symmetries that define the bulk ABM phase, $A_{\alpha
i}^{\text{ABM}}=d_{\alpha}\,\left(\hat{m}\pm i\hat{n}\right)_{i}$, are (i) \emph{chiral symmetry},
$\ZtwoChiral=\{E,\parityTwoD\cdot\time\}$, where $\parityTwoD$ is reflection in a plane containing
the chiral axis $\hat{l}=\hat{m}\times\hat{n}$, and (ii) \emph{gauge-orbit symmetry}, $\gaugeorbit$,
i.e. rotation by angle $\vartheta$ about the chiral axis, combined with a gauge transformation,
$e^{\pm i\vartheta}$, by phase angle $\mp\vartheta$.
A discrete analog of the ABM phase of bulk \He\ is obtained by breaking $\CfourVL$ rotational
symmetry, but restoring symmetry with appropriate elements from $\gauge$. In addition, $\time$
symmetry is broken, but chiral symmetry is present as invariance with respect to the combined
operation, $\time\cdot\Pi_{zx}$.
Thus, the discrete ABM phase is invariant with respect to the group obtained from these generators,
\ber
\hspace*{-4mm}
\CfourVLNT = \left\{E,e^{i\pi/2}C_4,e^{i\pi}C_4^2,e^{i3\pi/2}C_4^3,
             \time\Pi_{zx},e^{i\pi/2}\time\Pi_{zx'},
             e^{i\pi}\time\Pi_{zy},e^{i3\pi/2}\time\Pi_{zy'}\right\}
\,,
\eer
and is isomorphic to $\CfourVL$.
The full symmetry group is then
\be
{\mathsf H}_{\Asquare} = \twoDspin\times\ZtwoSpin\times\CfourVLNT
\,,
\ee
and the functional form of the discrete ABM phase is $A_{\alpha i}^{\pm} =
a(x,y)\,\hat{d}_{\alpha}\,\left(\hat{x} \pm i\,\hat{y}\right)_{i}$, where $a(x,y)$ is real an obeys
the $\CfourVL$ reflection symmetries in Eq. \ref{eq-C4v_reflections_az}.
The $\Asquare$-phase is not stable in the weak-coupling limit. However, several chiral phases are
found to be stationary points of the GL functional for strong-coupling values of the
$\beta$-parameters appropriate for high pressures. The phase diagram at high pressures will be
discussed in a separate report.

\section{Numerical Methods}

\begin{figure}[t]
\begin{center}
\includegraphics[width=1.0\linewidth,keepaspectratio]{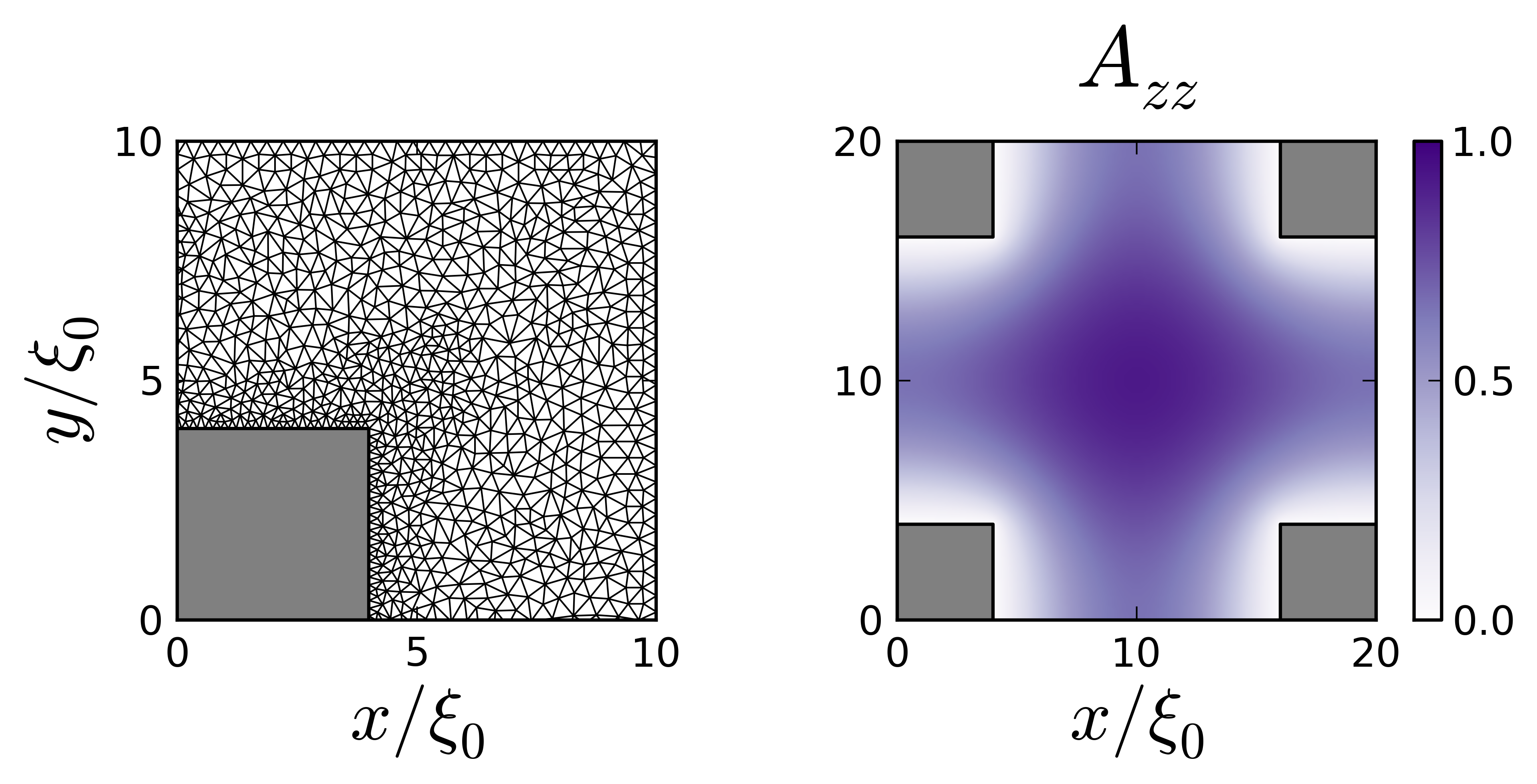}
\end{center}
\caption{(Color online) 
Left panel: One-fourth of the unit cell showing the triangular computational grid. 
The grey region defines the area occupied by the post.
Right panel: Order parameter amplitude, $A_{zz}(\vec{R})$, of the z-aligned Polar phase for 
$L=20\,\xi_{0}$, $d=8\,\xi_{0}$, $T=0.9T_{c}$, and $\hat{d}=\hat{z}$.
The order parameter is real and scaled in units of the bulk B-phase order parameter,
$\Delta_{\text{B}}(T)$.
}
\label{fig-Mesh+Polar_phase}
\end{figure}

To compute the order parameter which minimizes the GL functional we implement a finite element
method (FEM).\cite{zienkiewicz05} We discretize the \He\ unit cell with an unstructured triangular
mesh generated with the code \texttt{Triangle}.\cite{she96} This type of mesh permits spatially
varying triangular element sizes, which we use to provide finer spatial resolution in regions near
boundaries and sharp corners as shown in the left panel of Fig. \ref{fig-Mesh+Polar_phase}.
Also, an unstructured mesh does not enforce any point symmetry that a periodic mesh possesses. Thus,
the residual symmetries of the phases we find result from interaction terms in the GL functional
combined with pair-breaking and periodicity represented by the boundary conditions.

For the FEM we represent the order parameter with quadratic Lagrange interpolating functions defined
on each element.
The Lagrange interpolating functions are determined by the values of the order parameter at six
nodes corresponding to the vertices and midpoints of edges of each element. The order parameter
field defined at the nodes of each element is continuous across the entire domain.
The resulting integration over the domain then separates into independent integrals over each
element which we evaluate numerically with Gauss-Legendre quadrature.\cite{abramowitz70}

We minimize the discretized GL functional using an implementation of the conjugate gradient
algorithm, \texttt{CG\_DESCENT}.\cite{hag06}
The gradient, $\mathbf{G}[A]\equiv\delta\Omega/\delta A^{\dag}(\vec{R})$, is evaluated at each node
within the finite element scheme and input as the gradient in the conjugate gradient method.
We set convergence as $\mathrm{max}\left\{\left|\mathbf{G}_{i}[A]\right|\right\}<10^{-7}$, for all
$i$ degrees of freedom (i.e. all 9 complex components of $A$ at each node) which we determined to
yield no significant loss of accuracy compared to stricter tolerances.

\section{Stable Phases - Maximal Pair-breaking}

Figure \ref{fig-Mesh+Polar_phase} (right panel) shows the equilibrium order parameter for confined
\He\ with period $L=20\,\xi_{0}$ and post dimension $d=8\,\xi_{0}$ at temperature $T=0.9T_{c}$ for
the case of maximal pair-breaking by the interior boundary.
This is a spatially modulated z-aligned Polar ($P_z$) state in which only the z-orbital
component, $A_{zz}$, is non-vanishing. This phase breaks spin- and orbital rotation symmetry, but
preserves time-reversal symmetry.
Note that the polar amplitude is maximum in the center of the channel and decreases by approximately 
$50\,\%$ into both $x$- and $y$ channels.
The $P_z$ phase is an equal-spin pairing state and thus the more general representation for this
phase is $A_{\alpha i} = \Delta(\vec{R})\,\hat{d}_{\alpha}\,\hat{z}_{i}$, where $\hat{d}$ is a real
unit vector that defines the broken rotational symmetry in spin space. The $P_z$ phase with only
$A_{zz}\ne 0$ corresponds to $\hat{d}=\hat{z}$, and is degenerate with respect to the orientation of
$\hat{d}$ since we have neglected the nuclear dipole and Zeeman energies.
The $P_z$ phase belongs to the symmetry class of pairing states defined by the sub-group, ${\mathsf
H}_{P_z} = \twoDspin\times\ZtwoSpin\times\DfourhLpi\times\time$, as discussed in Sec.
\ref{Sec-Symmetry_Classes}

For the periods, $L\le 30\xi_{0}$, and temperatures within the region of stability of the $P_z$
phase, we find a finite Polar amplitude everywhere within the \He\ cavity, except at the post
boundaries. However for much larger periods, $L$, and the same channel width, $D=L-d$, the amplitude
of the $P_z$ order parameter appears to vanish deep within the $x$- and $y$ channels far from the
center, leaving a lattice of isolated islands of $P_z$ condensate in the center. This suggests there
may be a regime in which de-coupled $P_z$ condensates nucleate in the center region, but are not
phase coherent and do not exhibit superfluidity.

\begin{figure}[t]
\begin{center}
\includegraphics[width=0.95\linewidth,keepaspectratio]{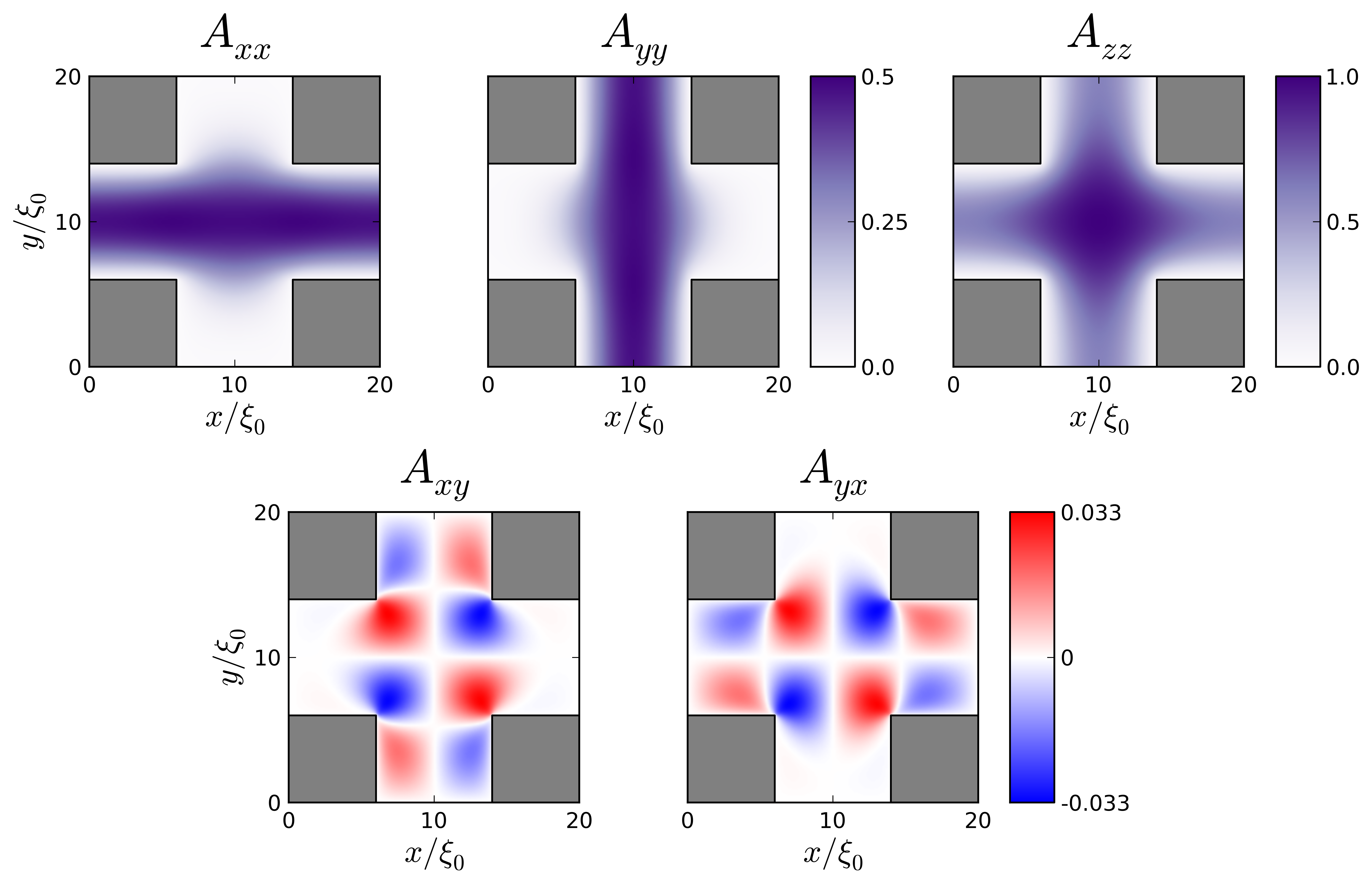}
\end{center}
\caption{(Color online) 
Order parameter components of the $\Bsquare$ phase plotted in the domain $V$ for 
$L=20\,\xi_{0}$, 
$d=12\,\xi_{0},$ and 
$T=0.7T_{c}$. 
All values are real and scaled by
the bulk B-phase order parameter $\Delta_{B}(T)$. 
Note the reduced scale of the $A_{xy}$ amplitude.}
\label{fig-B-like_phase}
\end{figure}

\begin{figure}[t]
\begin{center}
\includegraphics[width=0.8\linewidth,keepaspectratio]{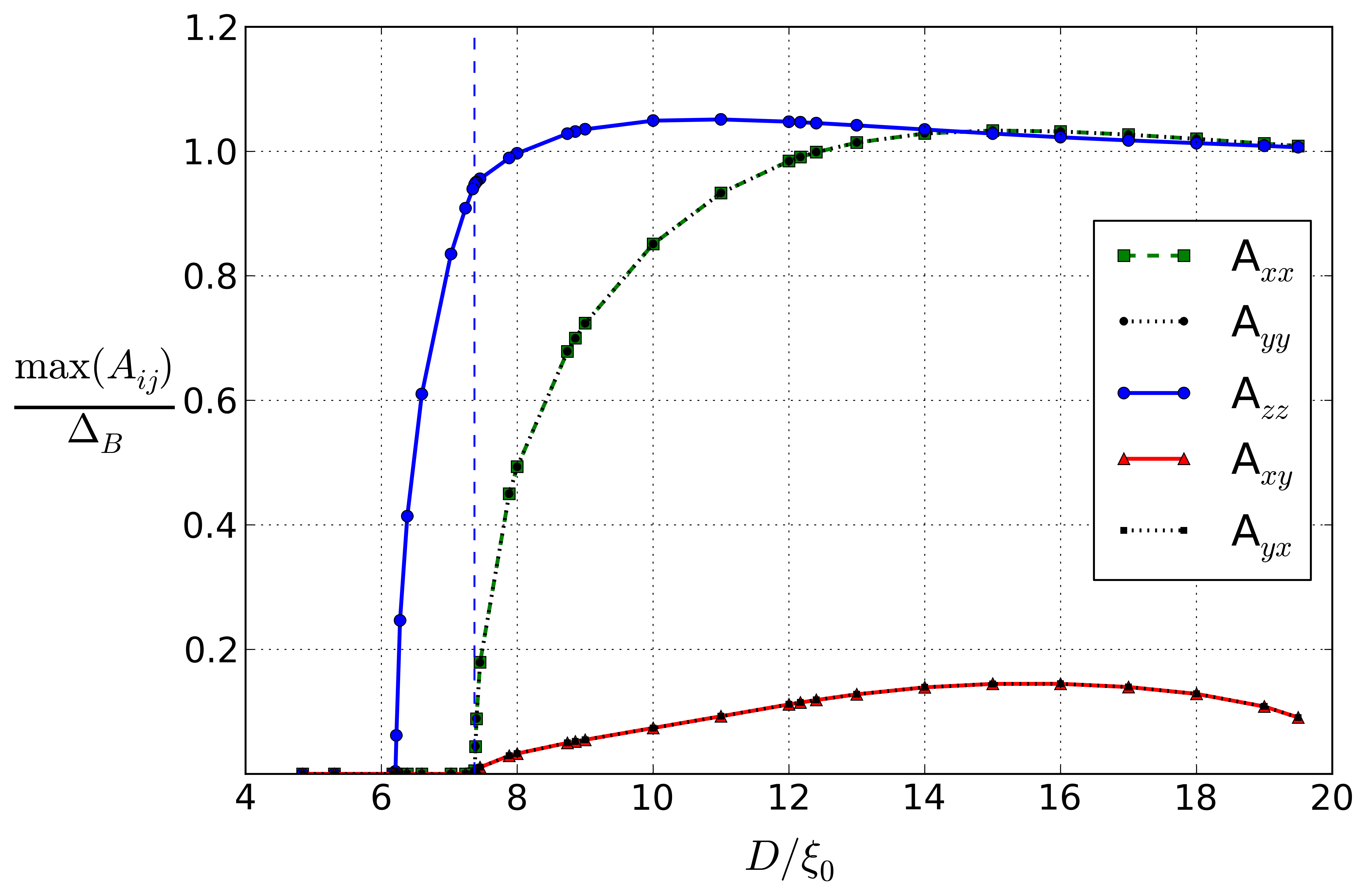}
\end{center}
\caption{(Color online) Order parameter amplitude as a function of the confinement length, $D$, for
$L=20\xi_{0}$ and $T=0.7T_{c}$. The amplitudes are taken as the maxima within the domain. Note
that maxima for $A_{xx}$ and $A_{yy}$ are equal, as are the maxima for $A_{xy}$ and $A_{yx}$, but 
suppressed compared to $A_{zz}$.
The dashed vertical line marks the $2^{\text{nd}}$ order $P_z$ to $\Bsquare$ phase transition.}
\label{fig-B_amplitudes}
\end{figure}

For the same period, post dimension and boundary conditions we also find a second stable phase in
the weak-coupling regime at a lower temperature. This phase ($\Bsquare$) also preserves
time-reversal symmetry, but has lower symmetry than that of the $P_z$ phase.
The $\Bsquare$ phase is similar to the bulk B-phase in that the order parameter is real, with
diagonal elements, $A_{xx}$, $A_{yy}$ and $A_{zz}$ in the center of the channel as shown in Fig.
\ref{fig-B-like_phase}.
However, the component $A_{xx}$ ($A_{yy}$) is strongly suppressed in the $y$-channel ($x$-channel),
and off-diagonal components, $A_{xy}$ and $A_{yx}$, appear at the corners of the posts.
It is also clear from Fig. \ref{fig-B-like_phase} that the components of the order parameter obey
the reflection symmetries: $A_{xx}(x,y)=A_{yy}(y,x)$ and $A_{xy}(x,y) = A_{yx}(y,x)$, and that the
diagonal components, $A_{xx}$, $A_{yy}$ and $A_{zz}$ are even functions of $x$ and $y$, while the
off-diagonal components, $A_{xy}$ and $A_{yx}$, are odd under $x\rightarrow -x$ or 
$y\rightarrow -y$. The remaining off-diagonal components are all zero: 
$A_{zx} = A_{xz} = A_{zy} = A_{yz} = 0$.
As discussed in Sec. \ref{Sec-Symmetry_Classes} these are the conditions imposed by the discrete 
sub-group, ${\mathsf H}_{\text{B}}=\DfourhLS\times\time$.
This is the maximal allowed point symmetry and is the discrete analog of the maximal subgroup
$\spinorbital$ for the bulk B-phase. Indeed, we recover the bulk B-phase for $L\rightarrow\infty$
and $d\rightarrow 0$, as indicated in Fig. \ref{fig-B_amplitudes}.
Note that the off-diagonal components are significantly smaller in magnitude than the diagonal
components and become negligible far from the post corners, except for $D\approx D_{c}(T)$, the
critical line separating the $P_z$ and $\Bsquare$ phases.

The phase transition from the $P_z$ to $\Bsquare$ phase is presented in Fig. \ref{fig-B_amplitudes},
which shows the maximal magnitudes for the components of the order parameter as a function of the
confinement length $D/\xi_0$ for fixed period, $L$, and temperature, $T$. The transition is
2$^{\text{nd}}$ order, i.e. continuous as a function of $D$ or $T$, with spontaneously broken
symmetry from ${\mathsf H}_{P_z}\rightarrow{\mathsf H}_{\Bsquare}$.
For confinement lengths onsetting at the critical value, $D_{c}(T)\rightarrow 4.1\xi_0$ at $T=0$ and
$L=20\,\xi_0$, the $x$ and $y$ components, $A_{xx}$, $A_{yy}$, $A_{xy}$ and $A_{yx}$, become finite,
signaling the transition to the $\Bsquare$ phase.
Close to the transition the $\Bsquare$ phase is locally a ``planar'' phase deep within the
channels due to the suppression of the orbital components normal to the boundary.
However in the central region the $\Bsquare$ phase is defined by all three diagonal components, as
well as the off-diagonal components, $A_{xy}$ and $A_{yx}$, allowed by $\DfourhLS$ symmetry.

\section{Weak-Coupling Phase Diagram}

\begin{figure}
\begin{center}
\includegraphics[width=0.95\linewidth,keepaspectratio]{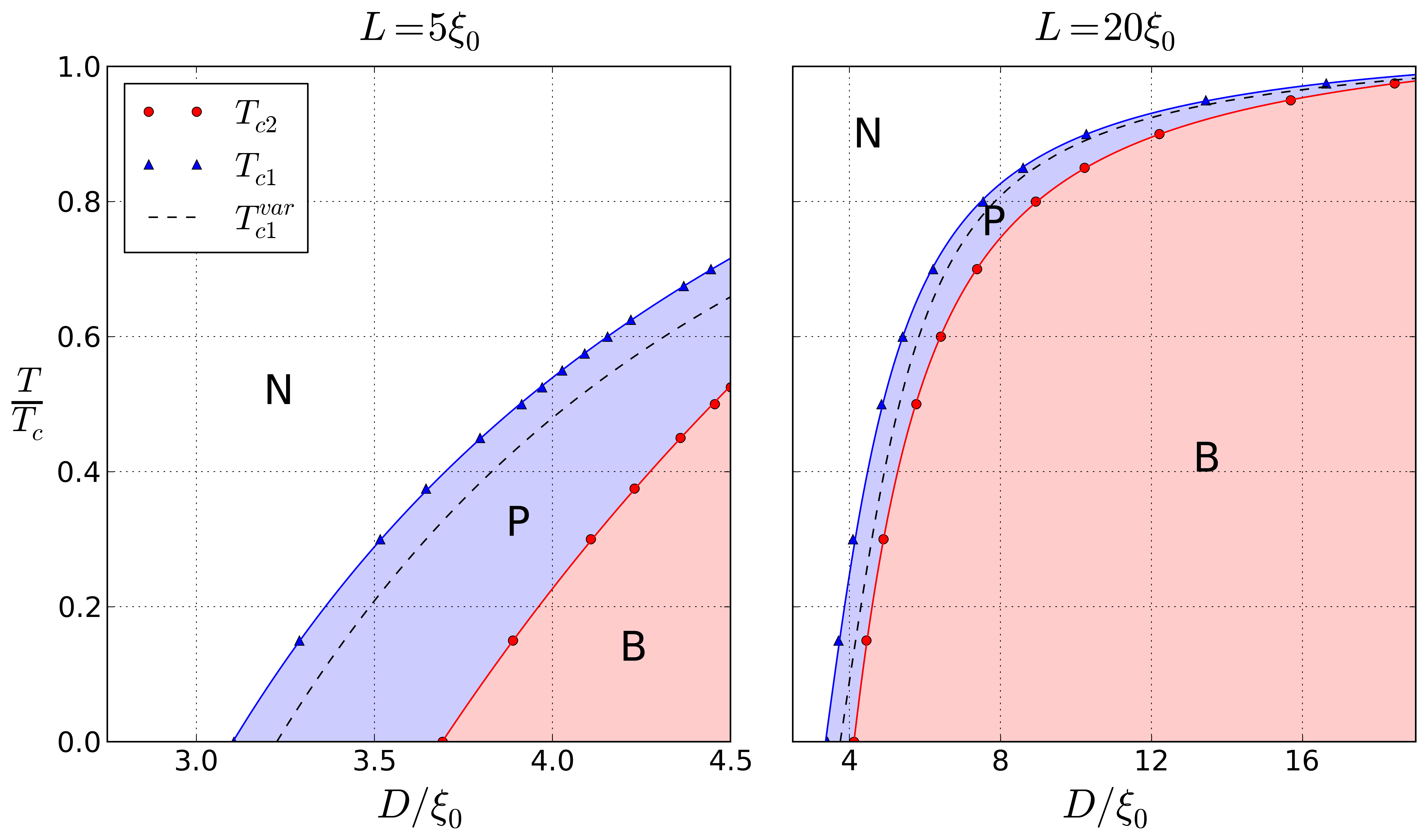}
\end{center}
\caption{(Color online) Phase diagrams for $L=5\xi_{0}$ and $L=20\xi_{0}$. The dashed curve is the
normal to $P_z$ transition obtained by the variational method. The solid curves are fits of the
transition data points to the functional form of the variational curve. Note that there is a range 
of confinement lengths, $D$, for which only the $P_z$ phase is realized.}
\label{fig-D-T_phasediagram}
\end{figure}

The phase diagram for superfluid \He\ in the weak-coupling limit as a function of reduced
temperature, $T/T_c$, and confinement length, $D/\xi_0$, is shown in Fig. \ref{fig-D-T_phasediagram}
for two values of the periodicity, $L = 5\xi_0$ (left panel) and $L = 20 \xi_0$ (right panel). These
two diagrams are qualitatively representative of the phase diagram for any $5 \le L/\xi_0 \le 30$.
In particular, we do not find any additional equilibrium phases as minima of the GL functional with
the weak-coupling material parameters.

The transition lines are found by classifying the phases based on non-negligible order parameter
components, and then bracketing the location of both normal to $P_z$ and the $P_z$ to $\Bsquare$
transitions. These brackets are refined until their width drops below a specified tolerance, which
we chose to be $0.025\xi_{0}$. Note that predicted phase boundaries are limited by the restriction
we place on the validity of the boundary condition for strong pair-breaking, i.e. $d\ge \xi_0/2$.

The phase boundary for the normal to $P_z$ transition is determined by a linear eigenvalue equation,
obtained by solving the linearized GL equation for the $P_z$ order parameter, $\alpha(T)\,A_{zz} -
K_{1}\left(\nabla_x^2 + \nabla_y^2\right)\,A_{zz} = 0$, within the domain $V$, and with boundary
condition, $A_{zz}\vert_{\partial V}=0$, for maximal pair-breaking. The eigenfunction, $A_{zz}\equiv
a_1(x,y)$, corresponding to the highest instability temperature, $T_{c_1}$, defines the spatial
profile of the first unstable mode of the $P_z$ phase. If we knew the exact functional form of the
first unstable mode, $a_1(x,y)$, we could obtain the phase boundary, $T_{c_1}(D,L)$, from the
equality in the Rayleigh-Ritz inequality,
\be
\alpha(T_{c_1}) \ge \frac{-\displaystyle\int_V d\vec{R}\left\{K_{1}|\nabla a(x,y) |^2\right\}}
                           {\displaystyle\int_V d\vec{R}\left\{|a(x,y) |^2\right\}}
\,.
\ee
In the absence of $a_1(x,y)$ we can obtain a lower bound on the $N$ to $P_z$ transition temperature
with a good approximation to the eigenfunction $a_1(x,y)$. Consider the following approximation to
the most unstable mode,
\begin{align}
\label{eq-approximate}
a(x,y)  & = \left[\frac{C(x) + C(y)}{1+C(x)C(y)}\right]\Theta(D/2-|x|) \Theta(D/2-|y|) 
\nonumber \\
		& + C(x) \Theta(D/2-|x|)\left[\Theta(-D/2-y)+\Theta(-D/2+y)\right]
\nonumber \\
		& + C(y) \Theta(D/2-|y|)\left[\Theta(-D/2-x)+\Theta(-D/2+x)\right]
\,,
\end{align}
where $C(x)=\cos(\pi x/D)$,
$\Theta(x)$ is the Heaviside step function, and $x$ and $y$ are defined on the domain $[-L/2,L/2]$. 
This function is piece-wise continuous at the interfaces between the central region and the $x$- and
$y$ channels, and satisfies the strong pair-breaking boundary condition, $a\vert_{\partial V}=0$. %
The variational result $T_{c_1}^{var}$, is shown in comparison to the exact numerical result for the
$N-P_z$ phase boundary, $T_{c_1}$ in Fig. \ref{fig-D-T_phasediagram}.

For the range of $L\le 30\xi_0$ that we consider, and for all $D\ge D_c(T)$ that gives a superfluid
transition, the $P_z$ phase is stable for a temperature range below $T_{c_1}$.
Furthermore, for a given $L$ there is a narrow range of confinement lengths, $D$, in which
only the $P_z$ phase is stable.
This is in sharp contrast to one-dimensional confinement in an infinite slab where the axial or
planar phases are stable under strong confinement.
The absence of these phases in the periodic confined geometry here is due to pair-breaking within
the two orthogonal $x$ and $y$ channels, the large cost in gradient energy for $x$- and $y$ orbital
components for strong confinement and the weak-coupling $\beta$-parameters.
Chiral phases, such as the $\Asquare$-phase and the chiral-$\CfourV$ phase will be discussed in a
separate report on GL theory of confined phases in the strong-coupling limit.

\section{Conclusions}

We have investigated the inhomogeneous phases of superfluid \He\ confined to a two-dimensional
lattice of square, sub-micron-scale boundaries (``posts'') with translational invariance in the
third dimension.
In the weak-coupling limit, and strong pair-breaking by the boundary post, we find an instability
from the normal state, at $T_{c_1} < T_{c}$ for bulk superfluid \He, to an equal-spin pairing state
with $z$-aligned Polar orbital order.
For fixed lattice spacing, $L$, there is a critical post dimension, $d_c$, above which
only the periodic polar phase is stable.
For $d < d_c$ we find a second, low-temperature phase onsetting at $T_{c_2} < T_{c_1}$ from the
polar phase to a periodic ``B-like'' phase. The low temperature phase is inhomogeneous, anisotropic
and preserves time-reversal symmetry, but unlike the bulk B-phase has only $\DfourhLS$ point
symmetry.
This or similar geometries may be realizable with current nano-fabrication processes, and could
therefore provide a potential avenue for experimental studies of the polar phase in \He\ in well
defined geometries.\cite{zhe13}
Further studies of \He\ in geometries with periodic confinement are expected to yield a large number
of tunable phases with unique broken symmetries and topological properties that are not realized in
bulk superfluid \He.

\begin{acknowledgements}
This research is supported by the National Science Foundation (Grant DMR-1106315).
We thank David Ferguson for many discussions and critique during the course of this work.
\end{acknowledgements}

%

\end{document}